\documentstyle[11pt,saltwshop,twoside]{article}
\begin{document}
\title{SALT and RR Lyrae Variables: Our Galaxy, The Magellanic Clouds 
and the Local Group}
\author{Michael Feast}
\affil{Astronomy Department, University of Cape Town, Rondebosch, 7701,
South Africa.\\
mwf@artemisia.ast.uct.ac.za}
\begin{abstract}
A review is given of the possibilities for the study of the kinematics
and metallicities of the old populations in galaxies of the Local Group
using SALT/PFIS observations of RR Lyrae variables.

\end{abstract}

\section{Introduction}
  In her summary talk at the recent Carnegie Cosmology conference,
Sandra Faber (2003) made two predictions:\\
 
1. ``The major era of chasing cosmological parameters is now closing"\\

2. ``Understanding galaxy formation.....will continue to occupy
cosmologists for some time......But there is no fundamentally new
physics \footnote{As distinct, presumably, from new and unusual
applications of known physics.} to be discovered there."\\

These are bold, perhaps rash, predictions, of the kind that are often falsified
by observations and SALT could well play a part in such falsification.

As regards galaxy formation, structure, and evolution, one gets some flavour
of the present situation by considering the dwarf spheroidal galaxies which are
often considered the simplest of systems. Below are summarized the 
conclusions of some papers, all on dwarf spheroidals, published in the last few 
months.\\
On the one hand:\\
Dwarf spheroidals are formed by the coalescing of clusters in dark matter
halos.\\
(astro-ph/0309202)\\
On the other:\\
Draco and UMi are weakly unbound systems - no significant dark matter.\\
(astro-ph/0309207)\\
But:\\
Draco is not the remnant of a tidally disrupted satellite but probably
is strongly dark matter dominated.\\
(ApJ 589, 798, 2003)\\
Though:\\
It is impossible to reproduce the tidal tail of UMi if it has a large
dark matter content.\\
(ApJ 586, L123, 2003)\\
But again:\\
UMi is one of the most dark matter dominated galaxies known.\\
(ApJ 588, L21, 2003)\\
And finally:\\
The extended outer structure of UMi could be extra-tidal stars either unbound
or within a dark matter halo.\\
(AJ 125, 1352, 2003)\\

I think it might be fair to summarize the above by saying that, not
only do we not know what dark matter is, we are not certain where it is,
and some would even say, if it is. It is worth noticing in passing that
it has recently been suggested
(Scarpa et al. 2003) that some globular clusters may contain dark matter
or that modified Newtonian dynamics is required.

As for Our Own Galaxy, it has been commonly supposed that its halo was formed 
by the infall of dwarf-spheroidal-like objects. It was also a
rather general belief that the halo of Our Galaxy is typical of the
halos of spirals.  However, we now know that the abundance of the $\alpha$
elements (i.e. [$\alpha$/Fe]) in dwarf spheroids is not similar to their
abundance in our halo 
(e.g. Shetrone et al. 2003, Tolstoy et al. 2003). Evidently infalling 
dwarf spheroidals are not
the major source of at least the inner halo of Our Galaxy. Furthermore,
recent work (Brown et al. 2003) suggests that the halo of M31 is distinctly 
different
from that of Our Galaxy. Thus bringing into question the notion of a 
``typical" halo.

It is clear that the current need is for detailed studies of the 
internal structure
and kinematics of all types of galaxies as a function of 
the chemical compositions
and ages of their component populations. 

\section{Variable stars and SALT}
In carrying out programmes of this type we need in each galaxy
to isolate homogeneous groups of objects. Variable stars are particularly
useful in this respect and amongst the variables there are three particularly
useful types. These are the Cepheids which trace young, metal rich populations
(the disc in Our Galaxy); RR Lyraes, old metal-poor objects,
which trace halo-type populations; and, the Miras which, as a function of
period, trace much of the intermediate age populations. These three types
of variable star can be used to study the distances, kinematics and 
distributions of different populations in Our Galaxy, in the Local Group
and, in some cases in more distant galaxies.

SALT is coming into operation in the era of massive variable star surveys.
This offers great opportunities and will doubtless have a major effect
on SALT programmes. 
MACHO, OGLE, MOA, SLOAN, QUEST are just some of the more
obvious current surveys. Such surveys will undoubtedly be extended in the
future either with the direct aim of finding variables or else
finding them as a by-product. For instance, planned surveys of the
sky for near-earth objects should also find variables in large numbers.
Not only are surveys finding very large numbers of variables to 
faint limits, but they are being found in systematic ways which is of great
importance in their use. In addition to this, surveys such as 2MASS allow
potential variables to be selected by their colours.

\section{The RR Lyrae Variables}
Of the three main types of variable star mentioned above, this paper will
concentrate on the RR Lyrae variables. It is in some ways quite appropriate
to discuss here the potential of SALT for studying these stars in our
own and other galaxies. Just over 50 years ago when the 1.9m telescope, now
at Sutherland and then in Pretoria, was the largest telescope in the 
southern hemisphere, David Thackeray and Adriaan Wesselink used it to
discover the first RR Lyrae variables in the Magellanic Clouds and Thackeray
used it to prove that the Hubble-Baade variables in the Sculptor
dwarf spheroidal were indeed RR Lyraes. These two discoveries were of major
importance. The first, increased the distance modulus of the Clouds by
1.5mag and was the most decisive evidence for a major increase in the
extragalactic distance scale. The second, was particularly important in
the context of the population scheme that Baade was then proposing.
The historical background to this work is given in the Baade-Thackeray
correspondence (Feast 2000).

RR Lyrae variables are found in globular clusters as well as the general field.
They are indicators of old, metal-poor populations, with [Fe/H] ranging
from about $-0.5$ to $-2.5$. Their pulsation periods range from
about 0.4 to 1.0 days if they are fundamental pulsators (the ``ab" type)
or about 0.2 to 0.5 days if they are overtone pulsators (``c" type).
Table 1 lists some estimates of their absolute magnitudes. 
\begin{table}
\caption{The Absolute Magnitude of RR Lyrae variables at $[Fe/H]= -1.5$}
\begin{tabular}{lcl}
Method & $M_{V}$ & reference\\
Parallax (HST) & +0.62 $\pm$ 0.16 & Benedict, et al. $2002^{1}$\\
Parallax (Hipparcos) & +0.40 $\pm$ 0.22 & Koen \& Laney 1998\\
Horizontal Branch & +0.63 $\pm$ 0.12 & Gratton 1998\\
Globular Clusters & +0.47 $\pm$ 0.12 & Carretta et al. 2000\\
$\delta$ Scuti variables & +0.49 $\pm$ 0.10 & McNamara 1997\\
Statistical parallaxes & +0.79 $\pm$ 0.13 & Gould \& Popowski 1998\\
      &      & \\
Adopted Mean      & +0.58 & \\
\end{tabular}

$^{1}$ see Feast (2002)
\end{table}

The adopted mean gives half weight to the result from Hipparcos
parallaxes because this has a large standard error. Some of the
other values may not be as well determined as their standard errors
might imply. For instance the statistical parallax solution depends on
a simple model of the galactic halo. Using the photometry of
RR Lyraes in the LMC (e.g. Clementini et al. 2003) with the adopted value of 
$M_{V}$ leads to an
LMC distance modulus of 18.53 in good agreement with  other values (see
e.g. Feast 2003). 

It has been known for some long while that the absolute magnitude of 
RR Lyrae variables depends on their metallicity. In its simplest
form this relationship can be written as:
\begin{equation}
M_{V} = \alpha [Fe/H] + \gamma
\end{equation}

The $M_{V}$ values given in Table 1 are for [Fe/H] = --1.5  assuming
$\alpha = 0.18$, a value often used. There has been considerable discussion
of the true value of $\alpha$. One reason that this is important is
that it affects the relative distances of globular clusters of
different metallicities when using RR Lyraes as the distance indicator.
These distances determine the absolute magnitudes at the main sequence
turn-off and thus the cluster ages. If $\alpha = 0.18$, metal-poor clusters
($\rm [Fe/H] \sim -2.2$) are older than metal-rich 
($\rm [Fe/H] \sim -0.7$) clusters
by about 3Gyr. If however, $\alpha = 0.39$ (Sandage 1993), then all
globular cluster would be about the same age, independent of metallicity
(see e.g., Sandage \& Cacciari 1990).
Thus, establishing the dependence of RR Lyrae absolute magnitudes on
metallicity is of important for understanding the early history and
evolution of Our Galaxy.

RR Lyrae variables are ideal spectroscopic targets for SALT. Photometry, 
from
large scale surveys is, or will be, available;
the short periods of the variables limit the
possible exposure time to less than 30min (or perhaps 60min for the
longer period stars), and their radial velocities and metallicities
can be adequately determined, for many purposes, at modest resolution
(say R $\sim$ 1000)
well within the range of PFIS. Thus they will be good targets for
the first generation of SALT instrumentation. RR Lyrae metallicities
are generally estimated using the method devised by Preston (1959).
This method compares the strengths of the Balmer lines with that of
the CaII(K) line, leading to a quantity $\Delta S$ which can be calibrated
as a function of [Fe/H] from high resolution observations of nearby
RR Lyraes. (For discussions of the calibration and accuracy of the method
see for instance: Suntzeff et al. 1991, Clementini et al. 1995,
Lambert et al. 1996, Solano et al. 1997, Fernley and Barnes 1997.) 

\section{RR Lyraes and the LMC}
Until recently the only spectroscopic metallicities known
for LMC RR Lyraes were of six stars observed with the ESO 3.6m telescope
at R$\sim 450$ and using the $\Delta S$ method
(Bragaglia et al. 2001). However, a short paper and a 
conference proceedings have now appeared giving some results from the
VLT. It is useful to look at these preliminary data in some detail as it
gives some idea of what may be achieved with SALT both in the Magellanic
Clouds and elsewhere.

Both these studies used the VLT with the FORS1 instrumentation. This
has a field of view of $7 \times 7$ arcmin, somewhat smaller than 
SALT/PFIS, and 19 slitlets. Note that FORS2 has the same size field but,
like PFIS, can accommodate more slits. Minniti et al. (2003) obtained
two exposures of 20 min on each of six fields in the LMC bar. There were
five to ten RR Lyraes per field
and the resolution was about 1000. Photometry 
(e.g. Clementini et al. 2003, Soszy\'{n}ski et al. 2003b) indicates
that these stars are at $V \sim 19.3$ and $B \sim 19.7$.
From these spectra Minniti et al. made an estimate of the velocity
dispersion. To do this they needed to correct the observed velocities
for pulsational effects (since they did not have full velocity curves).
This correction was made using a standard template with the phase
of the programme star
known from published photometry
\footnote{It is worth noting that in this type of work the template
used needs to be carefully chosen since strong and weak lines give
different velocity amplitudes (e.g. Oke et al. 1962).}
The velocity dispersion derived
then needed to be corrected for the scatter introduced
by the template method and also for the estimated uncertainty in the
radial velocity measurements. Table 2 shows their results.

\begin{table}
\caption{Velocity Dispersion of LMC RR Lyraes (Minniti et al. 2003)}
\begin{tabular}{cl}
Measured & 61 $\pm$ 7\\
Phase correction & 20\\
Measuring uncertainty & 22 \\
    & \\
``True" dispersion & 53 $\pm$ 10\\
    & $\rm km\,s^{-1}$\\
\end{tabular}
\end{table}

This result is of interest because the estimated dispersion is
larger than that of other objects in the LMC: i.e. 
young population, $\sim 9 \rm km\,s^{-1}$; Planetary nebulae, 
$\sim 20 \rm km\,s^{-1}$; Miras, $\sim 33 \rm km\,s^{-1}$. However, it is 
clearly
only a fore-taste of what might be done.

The other VLT/FORS1 study (Clementini 2003) summarizes the 
[Fe/H] results from Preston's $\Delta S$ 
(resolution about 800)
for about 100 RR Lyraes in the
LMC Bar. From a plot of $V_{o}$ against [Fe/H] 
Clementini and her co-workers find
that $\alpha$ in equation 1 above is 0.21. However, their figure 2 shows that 
this value is still rather uncertain, due mainly to the few points at
high and low metallicities
Considerably higher or lower values cannot at present be ruled out.
Most of the points lie in the range,
[Fe/H] --1.2 to --1.8. An estimate of the distribution of these points
together with the quoted uncertainty of a single value 
($\sigma_{[Fe/H]}$ = 0.2) suggests that much of the scatter in [Fe/H]
is observational. Evidence from galactic work (see e.g., Suntzeff et al. 1991) 
suggests that
it should be possible to derive relative [Fe/H] values by the $\Delta S$
method with an uncertainty of about 0.1 and this will be necessary to
study the distribution of RR Lyrae metallicities in the LMC in detail.

The two investigations just described suggest that with sufficient stars 
and careful (and probably repeated) observations, it will be possible
to study the kinematics and metallicities of RR Lyraes (and their relationship)
as a function of position in the LMC. With care it may even be possible
to study the kinematics and metallicities as a function of depth in the LMC.
$\Delta V_{o} =0.5 \rm mag$ corresponds to a depth in the line of
sight of $\sim$ 12 kpc at the distance of the LMC and for a spherical
subsystem this would be equivalent to a diameter of $\sim 14^{\circ}$
on the sky; a not unreasonable size for such a subsystem.

The current position regarding 
the discovery of RR Lyraes in the general field of the LMC
is as follows. The OGLE II survey (Soszy\'{n}ski et al. 2003b) which 
covered 4.5 sq.deg. over
the Bar region, has given data for 7600 RR Lyraes variables. This is an
average of about 26 variables per PFIS field. Numbers drop away from the
Bar. The MACHO survey covering about 10 sq.deg. gives data for 7900
RR Lyraes (Alcock et al. 1996). OGLE III now in progress covers about 
40 sq.deg. of the
LMC. In the outer field there will often only be one RR Lyrae in
an $8 \times 8$  arcmin SALT/PFIS field. It is clear that with the numbers of 
LMC RR Lyraes known, and surveys still in progress, the only limitations to 
doing a really good job on the kinematics and metallicities of this old
population will be the care taken in the work and the
amount of SALT/PFIS time available.

\section{RR Lyraes and the SMC}
So far as I am aware no spectroscopic studies have yet been published on
RR Lyraes in the SMC where these stars are about 0.5mag fainter than
in the LMC (i.e $V \sim 19.7 \rm mag$, $B \sim 20.0 \rm mag$). OGLE II gave 
data
on 571 RR Lyraes in a 2.4 sq.deg. field, or about 3 per SALT/PFIS field
(Soszy\'{n}ski et al. 2003a).
OGLE III covers an SMC area of about 15 sq. deg. It would be particularly
interesting to study the kinematics and metallicities of SMC RR Lyraes.
Young objects (e.g. Cepheids, Caldwell \& Coulson 1986) show the
SMC to be very extended in the line of sight, a depth to width ratio
of about 5 to 1. The estimated depth is between 15 and 20 kpc. Since
at the distance of the SMC, 17kpc corresponds to 
$\Delta V_{o} \sim0.6 \rm mag$,
it may be possible to resolve the depth structure in the RR Lyraes.

\section{RR Lyraes and the Dwarf Spheroidals}
Table 3 lists the dwarf spheroidal galaxies of the Milky Way subgroup
in order of their total visual absolute magnitude (taken mostly from 
van den Bergh 2000).  Also listed are estimates of their distance moduli,
the numbers of RR Lyraes currently known in each system (with references),
the estimated  approximate $B$ magnitude of these stars and the numbers
expected per SALT/PFIS field. Draco and UMi are of course too far north
for SALT (but could usefully be tackled by HET). Detailed studies of 
clearly old populations in the dwarf spheroidals which are generally
supposed to be relatively simple systems would be very valuable. 
It would of course be in parallel with the 
high resolution work, some already published
(e.g. Shetrone et al. 2003, Tolstoy et al. 2003), on the metallicities of 
RGB stars in these systems. 
Whilst the RGB stars are brighter
than the RR Lyraes, it seems difficult, at least
at present to be certain of the age of single stars of that type.
How much will be possible on the dwarf spheroids with SALT will
depend on how PFIS actually performs. We might well hope to be able to
study the kinematics and metallicities of RR Lyraes in Sextans, Sculptor
and Carina, and one might perhaps also hope to observe those in Fornax.
It is interesting to note that, so far as I am aware, there has been no
published spectroscopic study of the RR Lyraes in the core of the Sgr Dwarf
galaxy although these are relatively bright.

\begin{table}
\caption{The Dwarf Spheroidals of the Milky Way Sub-Group}

\begin{tabular}{lccrccc}
Galaxy & $M_{V}$ & Mod & N(RR) & Reference(RR) & $\sim$ B(RR) & PFIS \\
 & & & & & & \\
Sgr Dwarf & --13.8: & 17.0 & 2370 & 1 & 18.2 & 1 \\
Fornax    & --13.1  & 20.7 & 515  & 2 & 21.8 & 15 \\
LeoI      & --11.9  & 22.0 & 54   & 3 & 23.2 & 3  \\
LeoII     & --10.1  & 21.6 & 148  & 4 & 22.6 & 40 \\
Sculptor  & --9.8   & 19.7 & 226  & 5 & 21.1 & 50 \\
Sextans   & --9.5   & 19.7 & 36   & 6 & 20.9 & 6 \\
Carina    & --9.4   & 20.0 & 75   & 7 & 21.2 & 4 \\
Draco     & --8.4   & 19.5 & 263  & 8 &      &   \\
UMi       & --8.4   & 19.0 & 56   & 9 &      &   \\
\end{tabular}
\raggedright

References\\
1. Cseresnjes 2001 \\
2. Bessier \& Wood 2002 \\
3. Held et al. 2001 \\
4. Siegel \& Majewski 2000 \\
5. Kaluzny et al. 1995 \\
6. Mateo, Fischer \& Krzeminski 1995 \\
7. Dall'Ora et al. 2003 \\
8. see Dall'Ora et al. 2003 \\
9. Nemec et al. 1988\\
\end{table}

\section{RR Lyraes in Our Galaxy and the Sgr Dwarf Stream}
As regards Our Own Galaxy, some of the outstanding questions are as follows:\\
Was the Halo formed from infalling satellites?\\
If so, should we see more than the remnants of Sgr Dwarf?\\
Can other remnants be found kinematically?\\
Is the recently found ring at $\sim$ 20kpc from the centre 
(e.g. Ibata et al. 2003, Crane et al. 2003, Sikivie 2003, Helmi et al. 2003,
Rocha-Pinto et al. 2003, Martin et al. 2003) 
a satellite
remnant or a structural feature of the galactic disc?\\
Was the inner halo formed by monolithic collapse or by mixing of
infalling debris?\\
Are the RR Lyraes in the galactic Bulge different from those in the
solar neighbourhood or is the Bulge reddening law anomalous? (see, e.g.
Stuz, Popowski \& Gould 1999, Udalski 2003 )\\
What is the structure of the galactic bar and its relation to the Bulge?\\

The RR Lyraes in Our Galaxy are relatively thinly distributed over the sky.
Even in the main body of Sgr Dwarf there is unlikely to be more than one 
per SALT/PFIS field. They become somewhat more concentrated in the Bulge
where one expects $\sim$ 5 per SALT/PFIS field in the Baade window
around NGC6522 (Oort \& Plaut 1975).  However, there is no lack of 
galactic RR Lyraes. 
The SLOAN survey for instance has found 3000 in just 1000 sq.deg.
(Ivezi\u{c} 2003). The SLOAN RR Lyraes extend out to a distance of 
$\sim$100 kpc (i.e. to V magnitudes of 20-21), with clumping
probably connected
with the Sgr dwarf stream.

The QUEST survey 
(Vivas et al. 2001, Vivas, Zinn \& Gallart 2003, Zinn et al. 2003) is 
finding large numbers of RR Lyraes (probably
with considerable overlap with SLOAN). They report 498 in a 380 sq.deg.
region. They also report a clump at a distance of about 50kpc
(RR Lyraes with $V \sim 19.2 \rm mag$). For 16 of these they have two 
spectra each
(R = 800) from VLT/FORS1. The radial velocity error for a single measurement
is estimated as $\sim 20 \rm km\,s^{-1}$ and the velocity dispersion of their 
16 stars is $\sim 25 \rm km\,s^{-1}$. This is much lower than that expected for
halo objects and is consistent with the suggestion that these stars
are part of the Sgr dwarf stream. The mean [Fe/H] of these RR Lyraes
is --1.7.

There is  a great deal of work to do here which will clarify
our understanding of both the Sgr dwarf (and other) stream(s) and
the structure of the halo, including its very outer parts.
Currently there is much interest in whether the great-circle structure
of the Sgr dwarf stream implies a spherical halo potential, contrary
to the predictions of CDM
(e.g. Ibata et al. 2001, Majewski et al. 2003, Helmi 2003 and a large
number of papers on the Sgr dwarf stream generally). 
Evidently detailed work on the radial
velocities and metallicities of RR Lyraes in the Sgr dwarf stream (and
also its core) is of importance both for studies of the structure
and evolution of Our Galaxy and also, more generally, for the nature
and distribution of dark matter. 

\section{Conclusion}
It is clear that SALT/PFIS will make possible the detailed study of the
kinematics and metallicities of RR Lyraes in our own galaxy, in the
Magellanic Clouds and in Local Group galaxies. This should lead to a 
major advance in our understanding of the structure, dynamics, evolution
and origin of galaxies. To do this properly will require a major
effort but the likely rewards are very considerable. And, if in the process
new physics is revealed, so much the better.

\section{Acknowledgments}
 I am grateful to Dr T.D. Kinman for his comments on an initial draft
of this paper and for drawing my attention to some important references.

\end{document}